# Investigating the biological potential of galactic cosmic ray-induced radiation-driven chemical disequilibrium in the Martian subsurface environment

Dimitra Atri
*Center for Space Science*
*New York University Abu Dhabi*
*PO Box 129188, Saadiyat Island, Abu Dhabi, UAE*
Email: atri@nyu.edu

There is growing evidence suggesting the presence of aqueous environment on ancient Mars, raising the question of the possibility of life in such an environment. Subsequently, with the erosion of the Martian atmosphere resulting in drastic changes in its climate, surface water disappeared, shrinking habitable spaces on the planet, with only a limited amount of water remaining near the surface in form of brines and water-ice deposits. Life, if it ever existed, would have had to adapt to harsh modern conditions, which includes low temperatures and surface pressure, and high radiation dose. Presently, there is no evidence of any biological activity on the planet's surface, however, the subsurface environment, which is yet to be explored, is less harsh, has traces of water in form of water-ice and brines, and undergoes radiation-driven redox chemistry. I hypothesize that Galactic Cosmic Ray (GCR)-induced radiation-driven chemical disequilibrium can be used for metabolic energy by extant life, and host organisms using mechanisms seen in similar chemical and radiation environments on Earth. I propose a GCR-induced radiolytic zone, and discuss the prospects of finding such life with Rosalind Franklin rover of the ExoMars mission.

1. **Introduction**

The Viking mission, which included two landers, Viking 1 (1976-1982) and Viking 2 (1976-1980), was tasked with detecting possible biosignatures on the Martian surface. Although the mission generated excitement initially, it failed to detect any biological activity from samples collected from the surface. Although, recently upon reexamination of results from the Label Release experiment it has been argued that the possibility of extant life cannot be ruled out based on Viking results[1]. With our current understanding of the planet, a high level of radiation dose prohibits the survival of any stable ecosystem on the planet's surface. Based on studies of geological features over the years, it is generally believed that Mars had abundant liquid water, and habitable conditions could have existed during the early to middle Noachian (4.1 to 3.7 billion years ago). Dramatic climate change resulted in drying of water bodies, and eventually shrinking of habitable spaces available on the planet. On present day Mars, there is evidence of only trace amounts of water in the form of brines, polar caps, hydrated minerals, and large deposits of water ice in the shallow subsurface environment[2]. The chemical environment also seems hostile to life. Oxidant species and perchlorates have been detected on the surface, and if heat activated, have the capability to destroy potential biomolecules and chemical biosignatures.

Although the Martian subsurface environment is yet to be directly probed, measurements from surface rovers, and satellite observations have provided reliable data to characterize it. Satellite observations of exposed water ice provides evidence of the existence of water below the surface. Ice deposits occur at a depth as shallow as 1 meter below the surface, extending down to several kilometers in depth [2]. The Martian surface is bombarded by Galactic Cosmic Rays (GCRs), which are energetic charged particles and are able to penetrate a few meters below the surface[3]. Theoretical modeling of radiation propagation informs us about the energy deposition rate below the surface[3-5] As discussed later, this continuous supply of energy leads to the formation of a number of chemical species, some of which may be potentially useful for metabolic chemistry. This is an additional source of energy along with radionuclides present in the Martian regolith.

Based on the results of radiation chemistry experiments, and observations of interstellar ices and comets, it is highly likely that prebiotic molecules are present below the surface. The degradation of potential subsurface organic molecules, resulting from GCR penetration occurs on timescales of millions of years[4,5]. This degradation effect is therefore negligible if an active ecosystem currently exists on Mars, since microbes have much smaller cell turnover times. Although, we do not understand how chemistry transforms itself into biology, we do understand basic properties of life, and minimum physicochemical environmental conditions required for life as we know it. The main aim of the manuscript is to estimate the likelihood of extremophile survivability in the Martian subsurface environment, which is planned to be explored with Rosalind Franklin rover of the ExoMars mission[6]. One of the objectives of the ExoMars mission is to search for biomarkers; find evidence of present or extinct life on the planet[6]. In this manuscript, I investigate the plausibility of radiation-induced chemical disequilibrium as a source of metabolic energy for potential Martian biota. In the following section, I will investigate how the Martian subsurface environment can be suitable for life, and whether it can be detected with Rosalind Franklin (ExoMars) rover.

2. **Subsurface Environment**

For life to potentially survive on Mars, five basic requirements have to be satisfied: (1) Presence of organics to supply biological hardware, (2) a source chemical disequilibrium to supply energy for metabolic activity, (3) a medium for nutrient transport, such as water, (4) the ability to transform "harmful" chemical species to benign products, and (5) a radiation damage repair mechanism. I focus on the top 2 meters of the subsurface environment, which is also the subject of a planned investigation by Rosalind Franklin rover[6] (ExoMars mission). Since there is no in-situ measurement of the subsurface region, estimates can only be made based on a combination of numerical modeling of radiation penetration, laboratory radiation chemistry experiments, data from satellites and rovers. The subsurface radiation environment is dominated by GCR-induced secondary particles[3-5]. The interaction between GCR-induced radiation with the Martian soil is the main driver of chemistry in the region. Mars is extremely dry, but small pockets of water-ice, hydrated salts, and other chemicals (discussed later) upon interaction with radiation could potentially be able to provide chemical energy to sustain independent ecosystems.

2.1 Numerical modeling: We use the GEANT4 numerical model[7] to calculate the energy deposition rate in the Martian subsurface. The model simulates charged particle propagation with matter, and can be used to model particle interactions in planetary atmospheres and surfaces. GEANT4 simulations have been found consistent with measurements by the RAD instrument[8] on board MSL on the Martian surface. We follow the GEANT4 configuration described in Matthia et al. and calculate GCR propagation in the Martian atmosphere and regolith. GCR spectrum was obtained from the BON10 model[9] which consists of 87% protons, 12% alpha particles, and 1% Iron nuclei which acts as a substitute for heavier particles. The Martian atmospheric properties were obtained from the Mars Climate Database (MCD)[10], which compiles data from a number of instruments and is used widely in the Mars community. We validated our numerical approach with measurements of the background flux of GCRs made by RAD. We obtained a background dose rate of 0.59 mSv/day, which is consistent with 0.64±12 mSv/day[11] within instrumental uncertainties of RAD. Solar Proton Events (SPEs) can temporarily enhance the subsurface energy flux for a period of up to several hours in the top layer of the Martian subsurface. The energy of typical SPE protons entering the Martian atmosphere is a few 100 MeV, which can go up to 10 GeV in extreme cases[12]. Pavlov et al. (2006)[5] calculated the subsurface energy deposition using the averaged proton flux of 33 protons $cm^{-2}$ $s^{-1}$ and found the top 20 cm layer impacted by SPEs and found it to be $2\times10^8$ $eVg^{-1}s^{-1}$ on an average. However, a typical event of a total proton fluence of $10^9$ protons $cm^{-2}$ on Earth would translate to a flux of $\sim5\times10^3$ protons $cm^{-2}$ $s^{-1}$ on Mars (SPE assumed to be lasting for 24 hrs[13]), or about 150 times the average flux of 33 protons $cm^{-2}$ $s^{-1}$. This would result in a temporary enhancement in radiation dose over the background dose from GCRs in the top 20 cm of the subsurface, and might be damaging to potential lifeforms.

The energy deposition rate obtained from numerical modeling ranges between $10^5$ - $10^7$ $eVg^{-1}s^{-1}$ in the top 2 m depth of the Martian subsurface as shown in Figure 1. For comparison, deep subsurface ecosystems on Earth survive on $\sim 10^6$ $eVg^{-1}s^{-1}$ [14] which, as shown in Figure 1 exists at a depth of 1-2 meters. Geological evidence suggests a thicker Martian atmosphere in the past, which would reduce the GCR-induced energy deposition in the subsurface, making the proposed mechanism less relevant. GCR flux is also anti-correlated with solar wind activity, and a higher wind strength in the ancient past[15] would have resulted in a lower GCR flux. A combination of these two factors makes this mechanism relevant on Mars later during its evolution into a planet with a thin atmosphere.

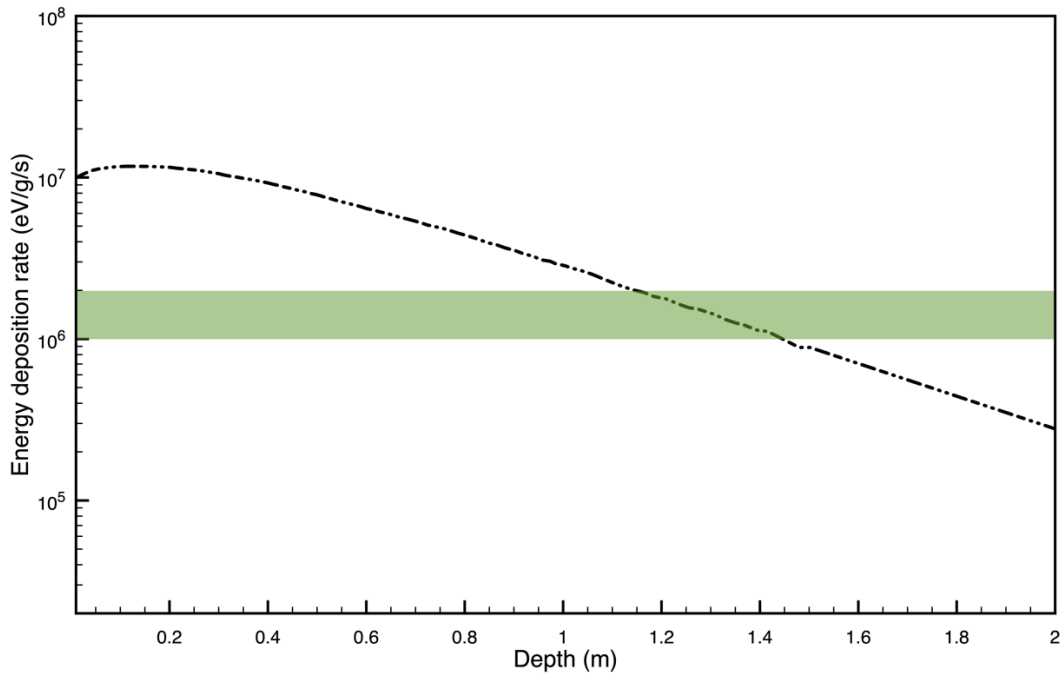

Figure 1: GCR-induced energy deposition rate below the surface of Mars based on the GEANT4 numerical model. The peak energy deposition rate is ~ $2\times10^7$ eVg$^{-1}$s$^{-1}$, which falls down to ~ $3\times10^5$ eVg$^{-1}$s$^{-1}$ at 2 m depth. It can be seen that a dose rate of $10^6$ eVg$^{-1}$s$^{-1}$ exists between 1-2 m depth on Mars, which is the measured dose rate in deep subsurface ecosystems on Earth[14].

2.2 Subsurface Organics: The presence of organics on the Martian surface has been established based on data from robotic missions[16,17]. They are either delivered through meteorites, or photo- and radio-chemically produced in the atmosphere, or through geological sources[16]. Organics can make their way below the surface through gardening. It has also been proposed that they can be produced in-situ from charged particle-induced chemistry (charged particle-ice interaction), similar to the way they are produced in interstellar ices and comets[3]. Radiation chemistry produces low-energy secondary electrons, which may result in reaction pathways not possible in photochemistry[18]. Radiolysis can also yield the formation of a number of biological useful molecules, as reported in some experiments[18,19]. It can also assist in carbon and nitrogen fixation, as seen in other cases[20,21]. Charged particle-induced reactions can produce complex organics such as amino acid precursors[19], Complex Organic Molecules (COMs), which are organic molecules of six or more atoms, in cometary ice[22], and abiotic ribose synthesis in interstellar ice analog experiments[23].

One should expect similar chemical products to be detected in the top Martian subsurface with the Rosalind Franklin rover. Subsurface organics are shielded from high dose of surface UV, but are exposed to GCR-induced secondary particles. Their rate of degradation below the surface has been calculated using numerical models, which occurs over a timescale of millions of years[4,5]. However, organics can be preserved, if they are a part of a biological system since it will likely have a repair mechanism, and also, the degradation rate would be negligible on shorter timescales expected from biological processes.

Yield of organics can be calculated based on laboratory experiments studying organics production from proton irradiation in ice mixtures and Mars-analog conditions. Kobayashi et al. irradiated a mixture of $CO_2$, CO, $N_2$ and $H_2O$ and studied the formation of amino acid precursors, formaldehyde and HCN[24]. Amino acids were formed after hydrolysis of radiolytic products. The yield of glycine was found to be 0.02/100 eV, which at peak rate assuming 100% efficiency would yield a maximum of $1.26 \times 10^{11}$ molecules/g annually. Other experiments have shown the G-value of formaldehyde, $H_2CO_3$ to be 0.2/100 eV[25] which would yield $1.26 \times 10^{12}$ molecules/g annually. Miyakawa (2002) with a $CO_2$-$N_2$-$H_2O$ atmosphere found the yield of the Uracil to be $1.6 \times 10^{-7}$ molecules/eV [26] or $\sim 10^8$ molecules/g annually.

2.3 GCR-induced chemical disequilibrium: Chemoautotrophic organisms are known to thrive by harvesting energy from environments with chemical disequilibrium. Photosynthesis is a prime example of radiation-induced chemistry, where solar photons initiate a series of reactions to produce ATP, which is harvested for metabolic energy. In absence of solar photons, the Martian subsurface environment would be dominated by radiation-induced chemistry, initiated by GCR secondaries[3-5]. GCR-induced radiolysis, which is a combination of ionization, excitation and dissociation, would lead to a series of radiochemical reactions. Radiolysis generates radicals, which react with the regolith, organic molecules and with other radicals, producing both oxidants and reductants which can be vital for metabolic reactions. These reactions lead to the production of more complex and stable molecules such as $H_2$, $O_2$ and $H_2O_2$ with time[19,27]. In pockets with abundant $CO_2$ and water ice, the $CO_2$-$H_2O$ irradiation chemistry would produce several organic compounds such as $CH_4$, $H_2CO_3$, $H_2CO$, and $CH_3OH$ [18]. Low temperatures ensure slow redox reactions resulting in oxidized species coexisting with reduced species[28]. Potentially, life would be able to catalyze reactions, which otherwise are inhibited at such low temperatures[29]. Therefore, a combination of radiation-induced reactions, along with low temperatures would assist in maintaining disequilibrium, since the natural rate of reactions would be extremely low, and reactions can occur only with the help of catalysts for metabolic purposes[28]. It is plausible that radiation-induced chemical disequilibrium can be potentially utilized as a source of metabolic energy for potential extant Martian biota. Diffusion of gases such as $H_2$, $O_2$ and CO can also be utilized as energy sources.

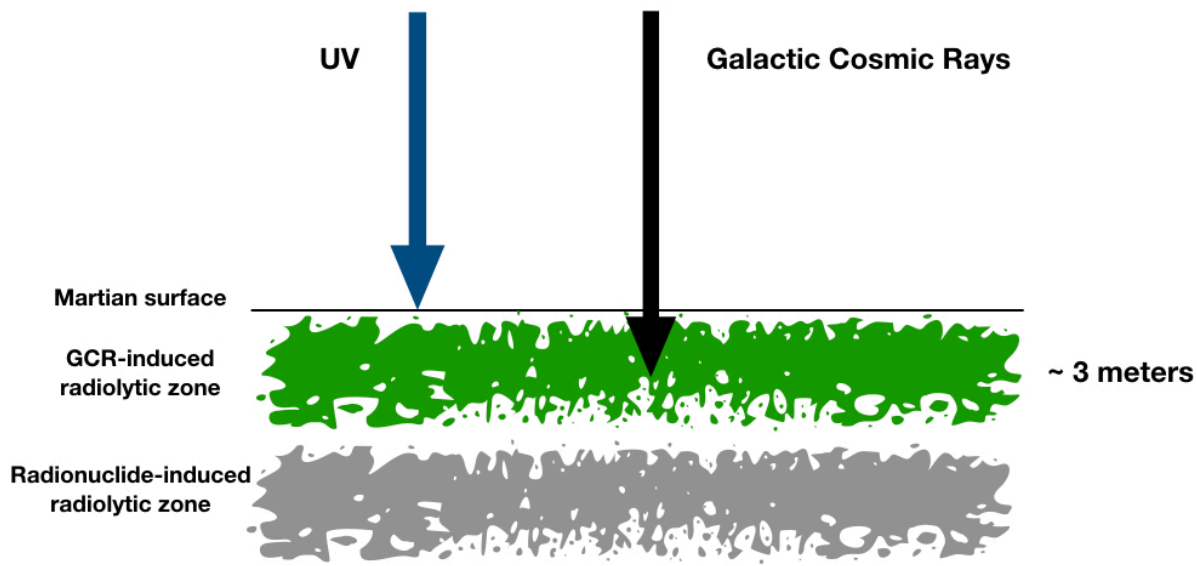

Figure 2: Proposed radiolytic habitable zones on Mars: Top 3 meters of the Martian subsurface is dominated by GCR-induced radiolysis, and below that with radionuclide-induced radiolysis[30]. The GCR-induced radiolytic zone has up to 2 orders of magnitude higher energy available for potential ecosystems than the radionuclide-induced radiolytic zone. The Martian surface is subjected to very high radiation levels, which is damaging to any potential ecosystem.

As shown in Figure 1, numerical models suggest that $10^6$ to $10^7$ $eVg^{-1}s^{-1}$ of energy is potentially available for redox chemistry[3]. However, not all energy will be utilized for metabolic activity, and this energy deposition rate represents the upper limit of energy availability. Next, we discuss possible mechanisms which can be used for metabolic activity.

a) GCR-induced $H_2$ production

One of the main modes of chemical energy is abiotic hydrogen production, which is vital for some deep-subsurface ecosystems that thrive independent of photosynthesis. Hydrogen is a great source of fuel where photosynthesis is not viable. Radiolysis from radioactive substances is known to produce hydrogen and has been known as a source of energy for subsurface life on Earth[14] and has been proposed as an energy source on Mars[30,31]. In deep-subsurface ecosystems, $H_2$ is used for both hydrocarbon synthesis, and as a source of metabolic energy[14,32,33]. $H_2$ is a source of energy for a number of electron acceptors and is important for Fischer-Tropsch (F-T) synthesis of organics. F-T synthesis is a reaction which catalytically converts CO and $H_2$ to hydrocarbons[33].

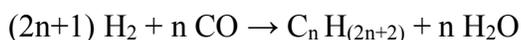

$$(2n+1)\ H_2 + n\ CO \rightarrow C_n H_{(2n+2)} + n\ H_2O$$

$H_2$ is also produced by a number of chemical processes such as serpentinization, which is hydrolysis of ferrous silicates. In addition to radioactive substances, I propose that subsurface ice would be exposed to GCR secondaries, where $H_2O$ gets dissociated into a number of products, ultimately producing $H_2$, $O_2$ and $H_2O_2$, and $H_2^{18}$ could be used for metabolic purposes[32]. Water-ice radiolysis is a well understood phenomenon, and $O_2$ and $H_2O_2$ are also found in the interstellar medium[18]. Methanogens are known to use $H_2$ and $CO_2$ to produce $CH_4$, both of which are available on Mars[34]. Methanogens are also found in permafrost, which are Earth-analogs to Martian subsurface. Methane had been detected at the Gale crater[35] and a source is yet to be identified but detection or lack thereof from the Martian orbit remains controversial[36]. One possible source of methane production could be from potential Martian methanogens in the subsurface using GCR-induced radiolysis products to power their metabolism. Metabolic energy can also be obtained by $H_2$ reacting with electron acceptors such as $Fe^{3+}$ or $SO_4^{2-}$.

b) GCR-induced sulfate reduction

*Desulforudis audaxviator* is known to survive in a radiation environment dominated by energetic alpha, beta and gamma radiation, away from solar photons, deriving metabolic energy from sulfate reduction, aided by the presence of radioactive rocks in a 2.8 km deep gold mine[37].

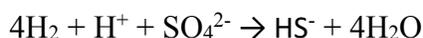

$$4H_2 + H^+ + SO_4^{2-} \rightarrow HS^- + 4H_2O$$

Radionuclide-induced sulfate reduction has been suggested as a potential source of metabolic energy on Europa[38]. I propose that sulfate reduction aided by GCR-induced radiation chemistry could be another possible mechanism in the Martian subsurface where the GCR flux is substantial (Figures 1 and 2). Aiding this mechanism is water ice, which is also available on Mars and discussed in detail later. The production rate of ATP molecules was estimated by multiplying the ATP yield[39] with the maximum energy deposition rate, which turns out to be ~ $1.9 \times 10^{14}$ molecules/g annually assuming 100% efficiency.

2.4 Other forms of disequilibrium: Chlorates and perchlorates are abundant on Mars. Perchlorates, if heat activated, can be harmful to organics, and some organisms. For example, they are found as water soluble contaminants that are harmful to human health. However, the presence of perchlorate does not exclude the possibility of life. Life, if evolved in this environment will develop capabilities to transform harmful chemicals to benign products. Some microbes have the capability to reduce chlorate and perchlorate, whose high reduction potential makes them efficient electron acceptors, and can be used for microbial metabolism[40]. Reduction of perchlorate to chlorate, chlorite and chloride has also been seen in some plants. One example of a perchlorate-reducing microbe is *Dechloromonas aromatica*, which in addition to perchlorate reduction, is also used for the bioremediative treatment of radionuclide contamination[40]. The key enzyme that enables the reaction is chlorite dismutase which converts $ClO^{2-}$ to $Cl^-$ and $O_2$ in presence of acetate which acts as a reducing agent. It is debatable whether or not microbes can use perchlorate reduction as an energy source to survive on Mars, but it is certain that the presence of perchlorates does not rule out the possibility of microbial communities powering their metabolism using alternate mechanisms.

2.5 Liquid medium: The presence of at least trace amounts of water is a basic requirement for life as we know it. The average Martian temperature is 210 K and the diurnal range is between 184 K and 242 K. The summer temperature can rise up to 293 K. Life is known to survive in temperatures down to 77 K. The presence of salts on Mars has been reported over the years, which lowers the freezing point of water and absorbs atmospheric moisture. Magnesium perchlorate, magnesium chlorate and sodium perchlorate have been detected on the surface. Hydrated calcium perchlorate has been detected on the Gale crater. Water's freezing point can be lowered to 40 K in the presence of sodium perchlorate and to 70 K in the presence of magnesium perchlorate. More recently, spectral evidence of flowing hydrated salts (Recurring Slope Linea or RSL) on the Martian surface was reported[41]. The detection of RSL supports the idea that briny solutions can flow on Mars although it is not an evidence of liquid water. It is plausible that life could be embedded in these briny solutions since halophilic organisms, such as *Salinibacter ruber* are known to thrive in such high salt environments. Some such microbes are known to survive in salt brines with concentrations up to 11%. Subsurface water-ice sheets have been observed up to 1-2 m below the surface and extend several tens of meters in depth[2]. An ice layer can also act as an insulator and can aid in maintaining more moderate temperatures below the surface.

2.6 Protection from potentially harmful chemicals: One of the fundamental properties of life is the ability to control its chemical environment. For example, a cell wall provides protection but also enables the exchange of chemicals through osmosis. On Mars, potential organisms can use similar mechanisms to maintain a favorable chemical environment. One possibility could be a habitat in porous rocks which can act as a protective shield but also allows for chemical exchange. This mechanism would enable organisms to survive in the presence of oxidants such as hydrogen peroxide, metal oxides such as $Fe_2O_3$ and $FeO_4^{2-}$[42], and perchlorates (if they are indeed dangerous to Martian biology). Low temperature would assist in maintaining disequilibrium in such conditions since the natural rate of reactions would be extremely low, and reactions can only occur with the help of catalysts for metabolic purposes.

2.7 Radiation damage repair mechanism: One of the most important requirements for any organism to survive in high radiation conditions is to be able to effectively repair radiation-induced damage. A number of extremophiles are known to have capabilities to survive in high radiation environments. *Deinococcus radiodurans* is able to withstand radiation exposure up to $10^5$ Sv, which is several orders of magnitude higher than the Martian subsurface[43]. *Desulforudis audaxviator* is able to thrive despite prolonged permanent exposure to radiation, likely by developing a damage repair mechanism which needs to be studied. Given a comparable dose environment in the 1-2 m depth in the Martian subsurface[3] as shown in Figure 1, it is highly plausible that extant life might have developed mechanisms to deal with radiation.

**3. Discussion**

In absence of a theory of the origin of life, it is not possible to determine whether life could have originated on ancient Mars when the environment was more hospitable to life as we know it. There is growing evidence that radiation-induced chemistry is capable of producing prebiotic compounds in a number of extraterrestrial scenarios and could have assisted with the potential origin of life on Mars[3,18,19]. However, with a thicker atmosphere of ancient Mars, the subsurface energy deposition from GCRs would be much lower making this mechanism less relevant. Later, when the planet lost much of its atmosphere, such potential ecosystem(s) can be sustained with radiation-driven chemical disequilibrium proposed in the paper (see below). So far, there is no known mechanism to understand how the formation of prebiotic compounds can lead to the origin of life. However, if life did exist on ancient Mars, there is a possibility of survival, which can be estimated based on our knowledge of extremophiles that are known to survive in comparably harsh environments on Earth. Such microbial life could have either originated on Mars, or transported from elsewhere, including from the Earth. Deep subsurface ecosystems are the only proxies for potentially extant subsurface Martian life. *Desulforudis audaxviator*, found in a 2.8 km deep South African mine is possibly the best such example. It is a sulfate reducing bacterium capable of thriving in a radioactive environment away from solar photons [37]. The bacterium is capable of using radiation-induced sulfate reduction for metabolic purposes and has developed damage repair mechanisms, which are crucial for survival in a high radiation environment[37]. As we have discussed, it has properties which are key to survival in the Martian subsurface. *Deinococcus radiodurans* is yet another extremophile, which is capable of withstanding radiation dose several orders of magnitude higher than that on Mars[44]. Some other well-known examples of extremophiles, among many, with properties required for survival in such an environment are *Dechloromonas aromatica*, which is a perchlorate-reducing bacterium[40], *Salinibacter ruber*, a halophilic bacterium, and *Methanosarcina barkeri*, a methanogen.

Here is a summary of GCR-induced processes in the Martian subsurface environment:
- A source of chemical disequilibrium
- Energy deposition rate ~ $10^6$ - $10^7$ eVg$^{-1}$s$^{-1}$
- Enhanced electron production
- Radiolytic production of $H_2$
- Radiolytic production of sulfates

The Rosalind Franklin rover (ExoMars) is well suited to detect extant microbial life proposed in the manuscript. It is a rover on board the ExoMars mission (ESA and Roscosmos), designed to detect biosignatures on Mars, scheduled to be launched in summer 2022. The most important tool on the rover is the Subsurface drill, which is capable of penetrating and obtaining samples up to 2 meters below the Martian surface[6]. The top 2m of Martian subsurface at the Oxia Planum a clay-laden landing site, which is thought to have had a standing body of water in the past would be investigated for its exobiological potential[6]. There are a number of instruments on board the rover designed to identify and characterize possible biosignatures, and signs of extinct or extant life as described in details by Vago et al. [6]. The Mars Organic Molecule Analyzer (MOMA) for the study of organic molecules and possible chemical biosignatures. The close-up imager (CLUPI) for high-resolution color images with a resolution of up to 8 μm/pixel. The MicroOmega instrument for imaging crushed sample material with a resolution of 20 μm/pixel. The ADRON neutron and gamma ray detector to measure the subsurface hydrogen content up to a depth of 1 m. The Raman Laser Spectrometer (RLS) for the detection of organic functional groups. These instruments would assist in detecting signs of possible microbial life and/or chemical biosignatures. If radiation-induced biology as proposed in the manuscript is detected, one might also consider the possibility of a GCR-induced radiolytic zone, which is a range of depth below the planetary surface at which radiation is optimum for biologically useful redox chemistry. As shown in Figure 2, up to ~ $10^7$ $eVg^{-1}s^{-1}$ of energy is available on the top 3 meters of Mars where biology would be able to utilize radiation-induced products for metabolic purposes while also being able to repair damage. The proposed mechanism would be applicable on planetary objects with a thin or no atmosphere at all, such as on Europa, Moon, comets etc. provided that the other key features required for life are satisfied, as discussed earlier, and should be detectable with future planetary missions.

## Acknowledgements

I thank the three reviewers for their feedback which helped in improving the manuscript. This work is supported by the New York University Abu Dhabi (NYUAD) Institute research grant G1502.

## References

1   Levin, G. V. & Straat, P. A. The Case for Extant Life on Mars and Its Possible Detection by the Viking Labeled Release Experiment. *Astrobiology* **16**, 798-810, doi:10.1089/ast.2015.1464 (2016).
2   Dundas, C. M. *et al.* Exposed subsurface ice sheets in the Martian mid-latitudes. *Science* **359**, 199-201 (2018).
3   Atri, D. On the possibility of galactic cosmic ray-induced radiolysis-powered life in subsurface environments in the Universe. *Journal of The Royal Society Interface* **13**, 20160459 (2016).

4	Dartnell, L. R., Desorgher, L., Ward, J. M. & Coates, A. J. Modelling the surface and subsurface Martian radiation environment: Implications for astrobiology. *Geophys Res Lett* **34**, doi:Artn L02207
10.1029/2006gl027494 (2007).
5	Pavlov, A. A., Vasilyev, G., Ostryakov, V. M., Pavlov, A. K. & Mahaffy, P. Degradation of the organic molecules in the shallow subsurface of Mars due to irradiation by cosmic rays. *Geophys Res Lett* **39**, doi:Artn L13202
10.1029/2012gl052166 (2012).
6	Vago, J. L. *et al.* Habitability on Early Mars and the Search for Biosignatures with the ExoMars Rover. *Astrobiology* **17**, 471-510, doi:10.1089/ast.2016.1533 (2017).
7	Agostinelli, S. *et al.* GEANT4—a simulation toolkit. *Nuclear instruments and methods in physics research section A: Accelerators, Spectrometers, Detectors and Associated Equipment* **506**, 250-303 (2003).
8	Hassler, D. M. *et al.* The radiation assessment detector (RAD) investigation. *Space science reviews* **170**, 503-558 (2012).
9	O'Neill, P. M. Badhwar–O'Neill 2010 galactic cosmic ray flux model—Revised. *IEEE Transactions on Nuclear Science* **57**, 3148-3153 (2010).
10	Millour, E. *et al.* in *European Planetary Science Congress*.  2015-2438.
11	Matthiä, D. *et al.* The Martian surface radiation environment–a comparison of models and MSL/RAD measurements. *Journal of Space Weather and Space Climate* **6**, A13 (2016).
12	Atri, D. Modelling stellar proton event-induced particle radiation dose on close-in exoplanets. *Monthly Notices of the Royal Astronomical Society: Letters* **465**, L34-L38 (2017).
13	Atri, D. Stellar Proton Event-induced surface radiation dose as a constraint on the habitability of terrestrial exoplanets. *Monthly Notices of the Royal Astronomical Society: Letters* **492**, L28-L33 (2020).
14	Lin, L. H. *et al.* Radiolytic H-2 in continental crust: Nuclear power for deep subsurface microbial communities. *Geochem Geophy Geosy* **6**, doi:Artn Q07003
10.1029/2004gc000907 (2005).
15	Wood, B. E., Müller, H.-R., Zank, G. P., Linsky, J. L. & Redfield, S. New mass-loss measurements from astrospheric Lyα absorption. *The Astrophysical Journal Letters* **628**, L143 (2005).
16	Freissinet, C. *et al.* Organic molecules in the Sheepbed Mudstone, Gale Crater, Mars. *J Geophys Res-Planet* **120**, 495-514, doi:10.1002/2014je004737 (2015).
17	Eigenbrode, J. L. *et al.* Organic matter preserved in 3-billion-year-old mudstones at Gale crater, Mars. *Science* **360**, 1096-1100, doi:10.1126/science.aas9185 (2018).
18	Arumainayagam, C. R. *et al.* Extraterrestrial prebiotic molecules: photochemistry vs. radiation chemistry of interstellar ices. *Chem Soc Rev* **48**, 2293-2314, doi:10.1039/c7cs00443e (2019).
19	Kobayashi, K. in *Astrobiology*   (ed Kakegawa T. Yamagishi A., Usui T.)  (Springer, Singapore, 2019).
20	Gao, K. S. *et al.* Solar UV radiation drives CO2 fixation in marine phytoplankton: A double-edged sword. *Plant Physiol* **144**, 54-59, doi:10.1104/pp.107.098491 (2007).
21	Bickley, R. I. & Vishwanathan, V. Photocatalytically Induced Fixation of Molecular Nitrogen by near Uv-Radiation. *Nature* **280**, 306-308, doi:DOI 10.1038/280306a0 (1979).
22	Garrod, R. T. Simulations of Ice Chemistry in Cometary Nuclei. *Astrophys J* **884**, doi:ARTN 69

10.3847/1538-4357/ab418e (2019).

23  Meinert, C. *et al.* Ribose and related sugars from ultraviolet irradiation of interstellar ice analogs. *Science* **352**, 208-212 (2016).

24  Kobayashi, K. *et al.* Possible complex organic compounds on Mars. *Advances in Space Research* **19**, 1067-1076 (1997).

25  Gerakines, P., Moore, M. H. & Hudson, R. L. Carbonic acid production in $H_2O$: $CO_2$ ices. UV photolysis vs. proton bombardment. *Astronomy and Astrophysics* **357**, 793-800 (2000).

26  Miyakawa, S., Yamanashi, H., Kobayashi, K., Cleaves, H. J. & Miller, S. L. Prebiotic synthesis from CO atmospheres: implications for the origins of life. *Proceedings of the National Academy of Sciences* **99**, 14628-14631 (2002).

27  Loeffler, M. J., Raut, U., Vidal, R. A., Baragiola, R. A. & Carlson, R. W. Synthesis of hydrogen peroxide in water ice by ion irradiation. *Icarus* **180**, 265-273, doi:10.1016/j.icarus.2005.08.001 (2006).

28  Zolotov, M. Y. & Shock, E. L. A model for low-temperature biogeochemistry of sulfur, carbon, and iron on Europa. *Journal of Geophysical Research: Planets* **109** (2004).

29  Clark, B. C. Surviving the limits to life at the surface of Mars. *Journal of Geophysical Research: Planets* **103**, 28545-28555 (1998).

30  Tarnas, J. *et al.* Radiolytic H2 production on Noachian Mars: Implications for habitability and atmospheric warming. *Earth and Planetary Science Letters* **502**, 133-145 (2018).

31  Dzaugis, M., Spivack, A. J. & D'Hondt, S. Radiolytic H2 Production in Martian Environments. *Astrobiology* **18**, 1137-1146, doi:10.1089/ast.2017.1654 (2018).

32  Blair, C. C., D'Hondt, S., Spivack, A. J. & Kingsley, R. H. Radiolytic hydrogen and microbial respiration in subsurface sediments. *Astrobiology* **7**, 951-970, doi:10.1089/ast.2007.0150 (2007).

33  Schulz, H. Short history and present trends of Fischer–Tropsch synthesis. *Applied Catalysis A: General* **186**, 3-12 (1999).

34  Onstott, T. C. *et al.* Martian CH4: sources, flux, and detection. *Astrobiology* **6**, 377-395 (2006).

35  Giuranna, M. *et al.* Independent confirmation of a methane spike on Mars and a source region east of Gale Crater. *Nature Geoscience* **12**, 326-332 (2019).

36  Korablev, O. *et al.* No detection of methane on Mars from early ExoMars Trace Gas Orbiter observations. *Nature* **568**, 517-520 (2019).

37  Chivian, D. *et al.* Environmental genomics reveals a single-species ecosystem deep within earth. *Science* **322**, 275-278, doi:10.1126/science.1155495 (2008).

38  Altair, T., de Avellar, M. G. B., Rodrigues, F. & Galante, D. Microbial habitability of Europa sustained by radioactive sources. *Sci Rep-Uk* **8**, doi:ARTN 260
10.1038/s41598-017-18470-z (2018).

39  Schulze-Makuch, D. & Irwin, L. N. *Life in the Universe*.  (Springer, 2004).

40  Coates, J. D. & Achenbach, L. A. Microbial perchlorate reduction: rocket-fuelled metabolism. *Nature Reviews Microbiology* **2**, 569 (2004).

41  Ojha, L. *et al.* Spectral evidence for hydrated salts in recurring slope lineae on Mars. *Nature Geoscience* **8**, 829 (2015).

42  Lasne, J. *et al.* Oxidants at the surface of Mars: a review in light of recent exploration results. *Astrobiology* **16**, 977-996 (2016).

43  Battista, J. R., Earl, A. M. & Park, M. J. Why is Deinococcus radiodurans so resistant to ionizing radiation? *Trends Microbiol* **7**, 362-365, doi:Doi 10.1016/S0966-842x(99)01566-8 (1999).

44   Daly, M. J. OPINION A new perspective on radiation resistance based on Deinococcus radiodurans. *Nature Reviews Microbiology* **7**, 237-245, doi:10.1038/nrmicro2073 (2009).